\documentclass[lettersize,journal]{IEEEtran}
\usepackage{amsmath,amsfonts}
\usepackage{array}
\usepackage{subfig}
\usepackage{stfloats}
\usepackage{url}
\usepackage{verbatim}
\usepackage{graphicx}
\usepackage{cite}
\usepackage{microtype}
\usepackage{doi}
\usepackage{float}
\usepackage{mathrsfs}
\usepackage{threeparttable}
\usepackage{booktabs}
\usepackage{multicol}
\usepackage{multirow}
\usepackage{tabto}
\usepackage[ruled]{algorithm2e}
\usepackage{algpseudocode}
\usepackage{makecell}
\usepackage{array}
\usepackage{diagbox}
\usepackage{adjustbox}
\usepackage{hyperref}
\usepackage{textcomp}
\usepackage{stackrel,amssymb}
\usepackage{xcolor}

\begin{document}
 \title{Key-Exchange Convolutional Auto-Encoder for Data Augmentation in Early Knee Osteoarthritis Detection}
 
\author{Zhe Wang, Aladine Chetouani, Mohamed Jarraya, Yung Hsin Chen, Yuhua Ru, Fang Chen, Fabian Bauer, Liping Zhang, Didier Hans$^*$, Rachid Jennane$^*$
\thanks{Zhe Wang, Yung Hsin Chen and Mohamed Jarraya are with Department of Radiology, Massachusetts General Hospital, Harvard Medical School, Boston, 02114, USA (e-mail: zwang78@mgh.harvard.edu; ychen4@mgh.harvard.edu; mjarraya@mgh.harvard.edu).}
\thanks{Aladine Chetouani is with L2TI Laboratory, University Sorbonne Paris Nord, Villetaneuse, 93430, France (e-mail: aladine.chetouani@univ-paris13.fr).}
\thanks{Yuhua Ru is with Jiangsu Institute of Hematology, The First Affiliated Hospital of Soochow University, Suzhou, 215006, China. (e-mail: ruyuhua@163.com).}
\thanks{Fang Chen is with department of Medical School, Henan University of Chinese Medicine, Zhengzhou, 450046, China. (e-mail: chenfangyxy@hactcm.edu.cn).}
\thanks{Fabian Bauer is with Division of Radiology, German Cancer Research Center, Heidelberg, 69120, Germany (e-mail: fabian.bauer@dkfz-heidelberg.de).}
\thanks{Liping Zhang is with Athinoula A. Martinos Centre for Biomedical Imaging, Massachusetts General Hospital, Harvard Medical School, Boston, 02114, USA. (e-mail: lzhang90@mgh.harvard.edu).}
\thanks{Didier Hans is with Nuclear Medicine Division, Geneva University Hospital, Geneva, 1205, Switzerland. (e-mail: didier.hans@chuv.ch).}
\thanks{Rachid Jennane is with IDP Institute, UMR CNRS 7013, University of Orleans, Orleans, 45067, France (e-mail: rachid.jennane@univ-orleans.fr).}}

\markboth{Journal of \LaTeX\ Class Files,~Vol.~14, No.~8, August~2021}%
{Shell \MakeLowercase{\textit{et al.}}: A Sample Article Using IEEEtran.cls for IEEE Journals}

\maketitle
\begin{abstract}
Knee Osteoarthritis (KOA) is a common musculoskeletal condition that significantly affects mobility and quality of life, particularly in elderly populations. However, training deep learning models for early KOA classification is often hampered by the limited availability of annotated medical datasets, owing to the high costs and labour-intensive nature of data labelling. Traditional data augmentation techniques, while useful, rely on simple transformations and fail to introduce sufficient diversity into the dataset. To address these challenges, we propose the Key-Exchange Convolutional Auto-Encoder (KECAE) as an innovative Artificial Intelligence (AI)-based data augmentation strategy for early KOA classification. Our model employs a convolutional autoencoder with a novel key-exchange mechanism that generates synthetic images by selectively exchanging key pathological features between X-ray images, which not only diversifies the dataset but also ensures the clinical validity of the augmented data. A hybrid loss function is introduced to supervise feature learning and reconstruction, integrating multiple components, including reconstruction, supervision, and feature separation losses. Experimental results demonstrate that the KECAE-generated data significantly improve the performance of KOA classification models, with accuracy gains of up to 1.98$\%$ across various standard and state-of-the-art architectures. Furthermore, a clinical validation study involving expert radiologists confirms the anatomical plausibility and diagnostic realism of the synthetic outputs. These findings highlight the potential of KECAE as a robust tool for augmenting medical datasets in early KOA detection. Our implementation is publicly available at \url{https://github.com/ZWang78/KECAE}.
\end{abstract}


\section{Introduction}
\IEEEPARstart{K}{nee} osteoarthritis (KOA) is a degenerative condition characterized by the deterioration and damage of articular cartilage, alterations at joint edges, and reactive hyperplasia of the subchondral bone \cite{kneeoa}. The onset of KOA is influenced by multiple factors, including age, weight, mechanical stress, and trauma, among others \cite{multi-factor}. Patients often endure severe pain and restricted mobility, which can significantly diminish their quality of life and elevate their risk for chronic conditions such as cardiovascular disease \cite{cardiovascular}. Despite its prevalence, the aetiology of KOA remains unclear, and there is currently no definitive cure \cite{notclear}. As such, early detection of KOA is crucial to enable timely behavioural interventions, such as weight loss, which can delay its onset and slow disease progression \cite{weightloss}. To evaluate KOA severity, Kellgren and Lawrence introduced the widely recognized Kellgren-Lawrence (KL) grading system \cite{KL}. As outlined in Table \ref{KL_grades}, this system classifies KOA into five grades based on the presence and severity of symptoms. Although the KL grading system is extensively used, it remains a semi-quantitative approach that relies heavily on the subjective perception and judgment of individual medical professionals, which can result in significant variability in the evaluation of the same knee X-ray image across different practitioners \cite{shamir}.

\begin{table}[htbp]
\centering
\caption{Description of the KL grading system}
\setlength{\tabcolsep}{2mm}
\begin{tabular}{lll}
\toprule
Grade &  Severity & Description\\
\midrule
KL-0 & none & none of osteoarthritis\\
KL-1 & doubtful & potential osteophytic lipping \\
KL-2 & minimal & certain osteophytes and potential JSN\\
KL-3 & moderate & moderate multiple osteophytes, certain JSN,\\
& & and some sclerosis\\
KL-4 & severe & large osteophytes, certain JSN, and severe sclerosis\\
\bottomrule
\end{tabular}
\label{KL_grades}
\end{table}

Deep learning, a cornerstone of artificial intelligence, has emerged as a promising approach for KOA classification \cite{tiuplin}\cite{zhe}\cite{zhe_ViT}\cite{chen}. However, the scarcity of reliable and annotated medical datasets poses a significant challenge for training robust models, largely due to the high costs associated with data labelling \cite{dataset}. Currently, deep learning networks heavily rely on data augmentation techniques to enhance generalization performance \cite{heavily_rely}. Traditional data augmentation methods, such as random rotation, are limited in their capacity to introduce significant diversity, as they are derived directly from the original data and do not alter its intrinsic structure, failing to substantially increase the diversity of training samples. To address these limitations, several approaches utilizing Auto-Encoders (AEs) \cite{AE} and Generative Adversarial Networks (GANs) \cite{GAN} have been proposed for generating synthetic data, providing an efficient and innovative solution to data augmentation. For instance, Khozeimeh et al. \cite{khozeimeh2021combining} integrated a Convolutional Neural Network (CNN) \cite{cnn} with an AE to predict survival outcomes for COVID-19 patients. Their study demonstrated the potential of AEs in enhancing medical image datasets in improving data quality for diagnosis and prognosis tasks. Similarly, Pesteie et al. \cite{pesteie2019adaptive} trained Variational Auto-Encoder (VAE) \cite{kingma2013auto} generated synthetic images for ultrasound spine images and brain MRI data augmentation, leading to notable improvements in model accuracy compared to conventional training approaches, demonstrating the effectiveness of AE-based models in improving image classification and tumour segmentation tasks. On the other hand, GAN-based methods have also shown promise in addressing dataset limitations. Tanaka et al. \cite{gan_3d} demonstrated that generating artificial training data using GANs effectively mitigates dataset imbalance. Their GAN-based model, evaluated on benchmark datasets, showed that a Decision Tree (DT) classifier \cite{DT} trained on GAN-generated data performed comparably to one trained on the original dataset. Similarly, Frid-Adar et al. \cite{frid_gan} employed Deep Convolutional GANs (DCGANs) \cite{DCGAN} to synthesize lesion plaques from liver CT scans across different categories, such as cysts, metastases, and hemangiomas. These synthetic samples significantly enhanced the performance of CNN classifiers trained on augmented datasets. Despite these advances, GAN-based models often fall short in generating high-quality images that meet the stringent requirements of medical diagnosis. For tasks demanding fine texture details, such as the early stage of KOA, GAN-generated images may lack sufficient detail and fidelity. Furthermore, the inherent challenges of GAN training, including difficulty in achieving convergence and reaching Nash equilibrium \cite{NE}, hinder the reproducibility and reliability of experimental results.

Inspired by the above studies, this paper proposes a novel model based on a Convolutional Auto-Encoder (CAE) to generate new and valid X-ray data for the early stages of KOA. The innovation of this study lies in the model's ability to exchange and combine key and non-key features from different inputs, namely, healthy and osteoarthritic knee X-ray images, to generate diverse and effective synthetic data, thereby expanding the dataset. Unlike traditional data augmentation techniques, our approach fundamentally alters the underlying structure of the data, leading to a more impactful and robust augmentation outcome.

\begin{figure*}[htbp]
\centering  %图片全局居中
\includegraphics[width=1\textwidth]{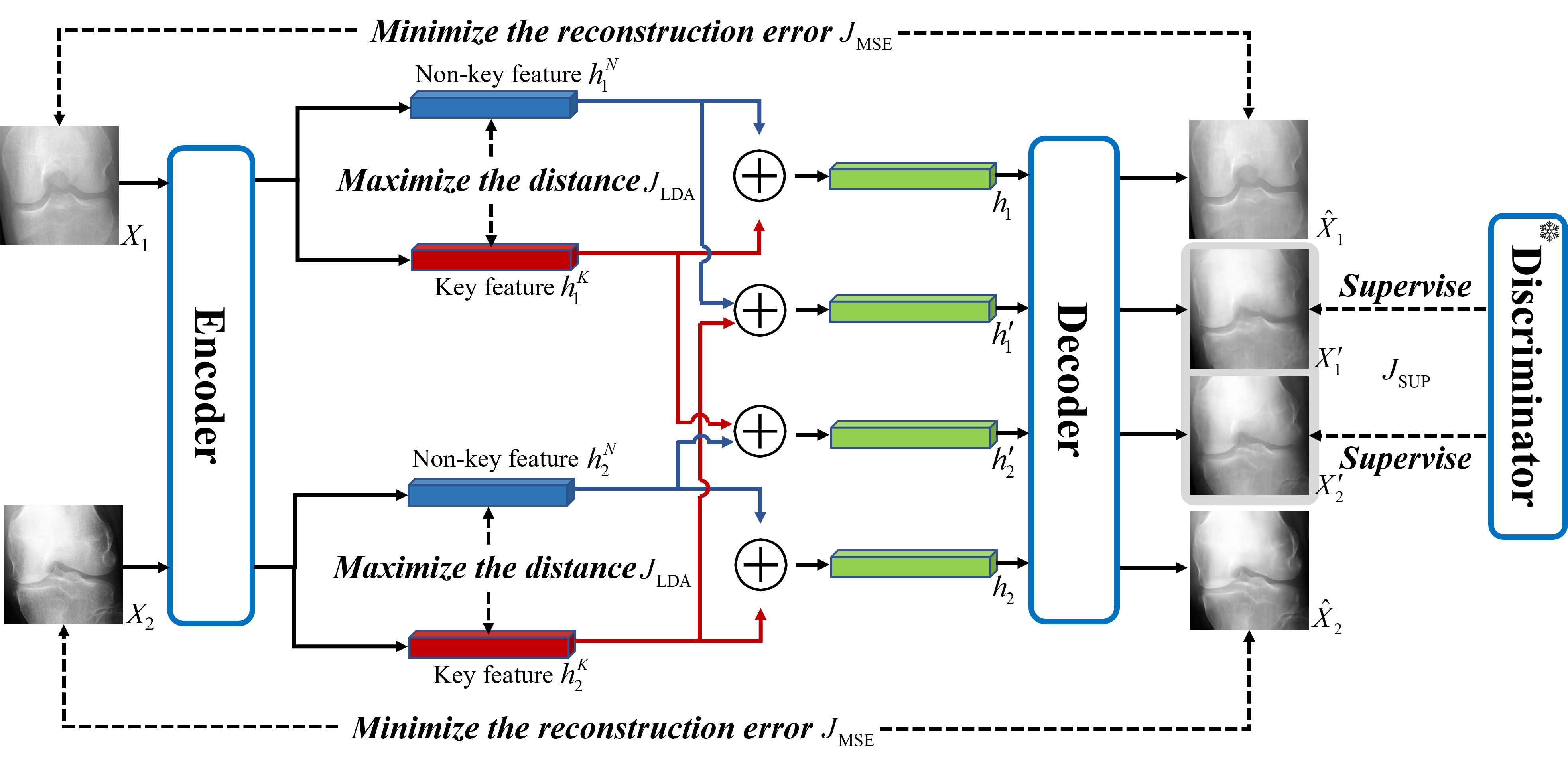}
\caption{The flowchart of the proposed KECAE model. Black, blue, and red arrows denote the overall data flow, non-key features, and key features, respectively. The symbol $\oplus$ represents element-wise addition. The encoder simultaneously processes two input images (a healthy knee X-ray and an osteoarthritic knee X-ray), generating two hidden vectors for each image: one representing key features and the other non-key features. These pairs of vectors are first combined through element-wise addition to produce two hidden vectors, which are then passed to the decoder to reconstruct the corresponding outputs. Simultaneously, the key feature vector of each input is exchanged and added element-wise with the non-key feature vector of the other input. These resultant vectors are then decoded to generate the key-exchanged outputs. To achieve this, we introduce a hybrid loss strategy comprising several components: the classical AE reconstruction loss, the supervision loss from a discriminator, and a distance loss implemented using the Fisher Linear Discriminant Analysis (LDA) algorithm \cite{fisher1}\cite{fisher2}. The details of this hybrid loss strategy are discussed further in Section \ref{hybrid_loss}.}
\label{flowchart}
\end{figure*}

The primary contributions of this study are summarized as follows:
\begin{itemize} 
\item[$\bullet$] A Key-Exchange Convolutional Auto-Encoder (KECAE) model is proposed to generate synthetic, clinically valid data for KOA classification. 
\item[$\bullet$] The generated images capture and exchange key and non-key features while preserving anatomical realism, enhancing dataset diversity. 
\item[$\bullet$] The augmented data significantly improve the performance of KOA classification models across various architectures. 
\item[$\bullet$] The clinical validity of the synthetic data is confirmed through expert radiologist evaluations, ensuring practical applicability. \end{itemize}

\section{Classical Auto-Encoder network}
\label{classical introduction}
\begin{figure}[htbp]
\centering
\includegraphics[width=0.45\textwidth]{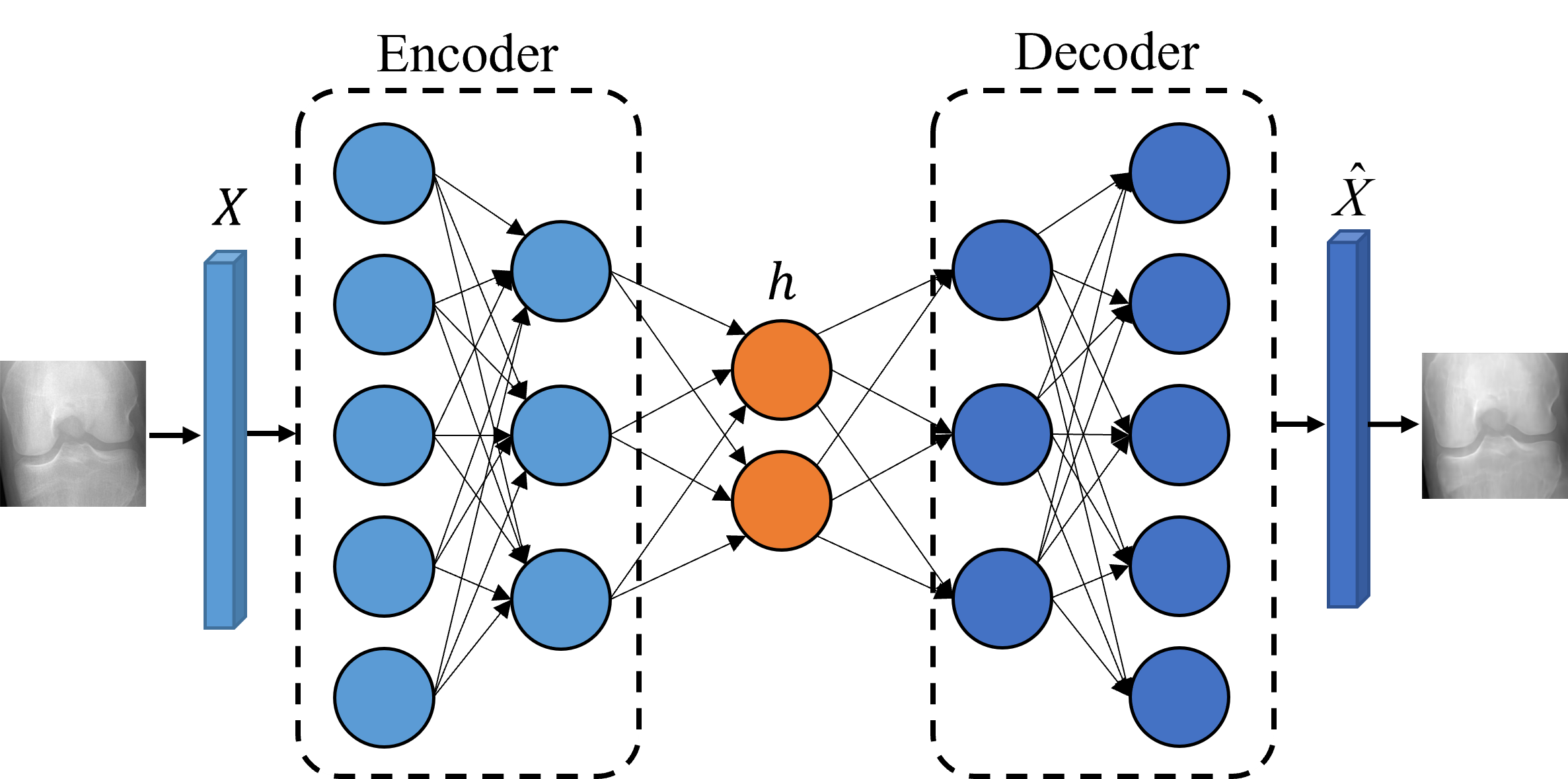}
\caption{The structure of the classical AE network.}
\label{AE}
\end{figure}

Prior to describing the proposed architecture, we provide a brief overview of the classical AE model. As shown in Fig. \ref{AE}, AE is a type of unsupervised learning model that comprises an encoder and a decoder. It utilizes the input data $X$ as its own supervision signal to enable the model to learn a mapping relationship and thereby obtains a reconstructed output $\hat{X}$. More precisely, the goal of the encoder is to convert the high-dimensional input $X$ into a lower-dimensional hidden variable $h$, with the aim of enabling the neural network to learn the most salient and informative features. The decoder restores the hidden variable $h$ of the hidden layer to the original dimension, and the goal of the model is to make its output a faithful approximate of the input (i.e., $minimize(dist(\hat{X}, X))$).

The encoding process of the input data $X$ to the hidden layers is as follows:

\begin{equation}
\label{encoder}
h = g_{\theta _1}(X) = \sigma(W_{e2}\sigma (W_{e1}X+b_{e1}) + b_{e2})
\end{equation}
where $g_{\theta _1}$ is the function of the encoder, $\sigma$ is an activate function, $W_{e1}$, $W_{e2}$ and $b_{e1}$, $b_{e2}$ represent weight and bias matrices of the encoder, respectively.

The decoding process of the hidden variable $h$ to the output layer is as follows:

\begin{equation}
\label{deoder}
%\hat{X} = g_{\theta _2}(h) = \sigma (W_2h+b_2)
\hat{X} = g_{\theta _2}(h) = \sigma(W_{d2}\sigma (W_{d1}h+b_{d1}) + b_{d2})
\end{equation}
where $g_{\theta _2}$ is the function of the decoder, $W_{d1}$, $W_{d2}$ and $b_{d1}$, $b_{d2}$ represent the weight and the bias matrices of the decoder, respectively.

Usually, the AE model minimizes the reconstruction error using the Mean Square Error (MSE) cost function, which is defined as:

\begin{equation}
\label{mse}
J_{\text{MSE}} = \frac{1}{N}\sum_{i=1}^{N}(\hat{X}-X)^2, \quad {\forall}X\in \mathcal{T}
\end{equation}
where $N$ is the amount of input data used during each training batch. $\mathcal{T}$ represents the training set.

%Compared to Principal Components Analysis (PCA) \cite{PCA}, the AE model can achieve non-linear dimensionality reduction of the data. The main advantage of the AE network is that it has a strong generalization capability and does not require data annotation, which is consistent with the conclusion of the AE-based work of \cite{reviewer_1}.

\section{Proposed method}
%In this section, the proposed learning model and the hybrid loss strategy will be presented, respectively.

\subsection{Learning model}
\label{proposed_approach}
As shown in Fig. \ref{flowchart}, our proposed model comprises three main components: an encoder, a decoder, and a discriminator. The encoder is based on two identical CNNs with non-shared parameters, where each CNN has 20 layers divided into seven blocks. Each block consists of a set of convolution layers of the same kernel of size 3$\times$3 with different depths (i.e., 32, 64, 128, 256, 512, 1024 and 2048) - a Batch Normalization (BN) layer - Leaky Rectified Linear Unit (LeakyReLU) layer with a predefined slope of 0.2. The structures of the encoder and the decoder are almost symmetrical. The decoder consists of seven DeConvolutional Neuron Network (DCNN) \cite{zeiler2010deconvolutional} blocks. These blocks consist of a set of deconvolutional layers of the same kernel size of 4$\times$4 with different depths (i.e., 2048, 1024, 512, 256, 128, 64, 32, and 1) - a BN layer - a LeakyReLU layer with a predefined slope of 0.2. In contrast to the classic AE model, our approach incorporates dual inputs ($X_1$ and $X_2$) and generates four hidden vectors: $h_1^N$, $h_1^K$, $h_2^N$, and $h_2^K$, where $h^N$ and $h^K$ represent the non-key and key feature vectors, respectively. Specifically, the encoder processes two input images ($X_1$ and $X_2$) simultaneously, producing two pairs of hidden vectors: ($h_1^K$ and $h_1^N$) for $X_1$, and ($h_2^K$ and $h_2^N$) for $X_2$. These vectors are then combined through element-wise addition to form two hidden vectors ($h_1$ and $h_2$), which are passed through the decoder to reconstruct the corresponding outputs ($\hat{X_1}$ and $\hat{X_2}$). Concurrently, the key feature vector $h^K$ from each input is exchanged and combined element-wise with the non-key vector $h^N$ from the other input. This process generates two new vectors ($h_1^\prime$ and $h_2^\prime$), which are then decoded to produce the key-exchanged outputs ($X_1^{\prime}$ and $X_2^{\prime}$). To extract and distinguish key and non-key features, we employed the pre-trained SOTA model \cite{wang2024transformer}, which has demonstrated the best performance in detecting early KOA stages (i.e., KL-0 and KL-2), to supervise the key-exchanged outputs ($X_1^{\prime}$ and $X_2^{\prime}$) by assessing whether their labels are correctly exchanged. Ultimately, the proposed KECAE model produces four outputs: $\hat{X_1}$, $X_1^{\prime}$, $\hat{X_2}$, and $X_2^{\prime}$. Among these, $X_1^{\prime}$ and $X_2^{\prime}$ are critical for data augmentation, as they represent the key-exchanged outputs that introduce diversity.

% \begin{figure}[htbp]
% \centering
% \subfloat[]{
% \begin{minipage}[t]{0.23\textwidth}
% \centering
% \includegraphics[width=1\textwidth]{encoder.png}
% \end{minipage}
% }
% \subfloat[]{
% \begin{minipage}[t]{0.23\textwidth}
% \centering
% \includegraphics[width=1\textwidth]{decoder.png}
% \end{minipage}
% }
% \caption{Structures of the encoder (a) and the decoder (b) in KE-CAE model.}
% \label{encoder_decoder}
% \end{figure}

%The structures of our encoder and decoder modules are shown in Fig.\ref{encoder_decoder}. 

\subsection{Hybrid loss strategy}
As presented in Section \ref{classical introduction}, MSE (Eq. \ref{mse}) is usually used as a reconstruction loss function. In this study, as there are two reconstruction outputs $\hat{X_1}$ and $\hat{X_2}$, two MSE losses $J_{\text{MSE}}(X_1, \hat{X_1})$ and $J_{\text{MSE}}(X_2, \hat{X_2})$ are thus obtained. The $J_{\text{MSE}}$ is defined as follows:

\begin{equation}
\label{MSE}
J_{\text{MSE}} = J_{\text{MSE}}(X_1, \hat{X_1}) + J_{\text{MSE}}(X_2, \hat{X_2})
\end{equation}

Two key-exchanged outputs $X_1^{\prime}$ and $X_2^{\prime}$ are supervised by the pre-trained discriminator to assess whether their labels are correctly exchanged. The Cross-Entropy (CE) losses, $J_{\text{CE}}(X_1^{\prime})$ and $J_{\text{CE}}( X_2^{\prime})$, are thus obtained, respectively. The supervision loss function, $J_{\text{SUP}}$, is defined here by:

\begin{equation}
\label{CE2}
J_{\text{SUP}} = J_{\text{CE}}(X_1^{\prime},Y_2) + J_{\text{CE}}( X_2^{\prime},Y_1)
\end{equation}
where $Y_1$ and $Y_2$ represent the real labels of $X_1$ and $X_2$, respectively.

In addition to $J_{\text{SUP}}$, to further separate the key and non-key features from each individual image, we employ the Fisher Linear Discriminant Analysis (LDA) algorithm \cite{fisher1}\cite{fisher2} to maximize these two feature vectors distance. It is computed as follows: 

\begin{equation}
\label{FDA}
J_{\text{LDA}} = \frac{({\sigma^{N}})^{2} + ({\sigma^{K}})^2}{\left | \mu^N- \mu^K \right |^{2}}
\end{equation}

\begin{equation}
\mu^i=\frac{1}{M}\sum_{m = 1}^{M}x_{m},\quad(\sigma^i)^2=\frac{1}{M}\sum_{m = 1}^{M}(x_{m}-\mu^i)^2
\end{equation}
where $\mu^i$, $(\sigma^i)^2$, $M$ denote the mean, variance, and number of elements ($x_{1}$, ..., $x_m$) in the feature vector, respectively. $i$ $\in$ $\left\{N, K\right\}$ represents the non-key $N$ and key feature $K$ vectors.

Finally, the hybrid loss function can be formulated as:
\begin{equation}
\label{hybrid_loss}
J_{\text{hybrid}} = J_{\text{MSE}} + \lambda_1 J_{\text{SUP}} + \lambda_2 J_{\text{LDA}}
\end{equation}
% mse的系数为1是为了保证ae的特征提取能力
where $\lambda_1$ and $\lambda_2$ are hyper-parameters, which will be discussed in Section \ref{hyperparameters}.

Overall, $J_{\text{hybrid}}$ achieves the global functionality of the proposed model by combining different loss functions. Specifically, $J_{\text{MSE}}$ serves as the foundational loss for the AE-based architecture and thus is given a weight of 1. On the other hand, $J_{\text{LDA}}$ is responsible for maximizing the distance between two extracted feature vectors to separate key and non-key features, while $J_{\text{SUP}}$ supervises whether the exchanged-key-feature images correspond to the new labels. Together, $J_{\text{LDA}}$ and $J_{\text{SUP}}$ facilitate the extraction of key and non-key features from an image, with $J_{\text{LDA}}$ focusing on feature separation and $J_{\text{SUP}}$ ensuring the quality of the separation. The impact of the hyper-parameters $\lambda_1$ and $\lambda_2$ on the model's performance will be analyzed and discussed in Section \ref{hyperparameters}. For greater clarity, Algorithm \ref{algorithm1} describes the training process of the proposed approach.

\begin{algorithm}
\caption{Algorithm of the proposed approach}
\textbf{Input:} $X_1,X_2$, and all initial parameters\\
\textbf{Output:} $\hat{X_1}, \hat{X_2}, X_1^\prime, X_2^\prime$, and learned parameters\\
\While{\textnormal{\textit{not converged}}}{
\textbf{Get} $h_1^N, h_1^K; h_2^K, h_2^N \leftarrow encoder(X_1, X_2)$ \Comment{Eq. \ref{encoder}}\\
\textbf{Get} $h_1 \leftarrow h_1^N \oplus h_1^K$, $h_2 \leftarrow h_2^N \oplus h_2^K$\\
\textbf{Get} $\hat{X_1}, \hat{X_2} \leftarrow decoder(h_1, h_2)$ \Comment{Eq. \ref{deoder}}\\
\textbf{Exchange} $h_1^K$ $\leftrightarrows$ $h_2^K$\\
\textbf{Get} $h_1^{\prime} \leftarrow h_1^N \oplus h_2^K$, $h_2^{\prime} \leftarrow h_2^N \oplus h_1^K$\\
\textbf{Get} $X_1^{\prime}, X_2^{\prime} \leftarrow decoder(h_1^\prime, h_2^\prime)$ \Comment{Eq. \ref{deoder}}\\
\textbf{Compute} $J_{\text{hybrid}}$  \Comment{Eq. \ref{hybrid_loss}}\\
\textbf{Update} Parameters of the encoder and decoder\\
}
\label{algorithm1}
\end{algorithm}

\section{Experimental settings}
It is noteworthy that, in the early stages of KOA, KL-2 signifies the presence of JSN, accompanied by visible osteophytes on the joint surface and along the edges of the knee bones. In contrast, KL-1 represents an earlier and subtler stage of KOA, where the indicators of the condition are less pronounced \cite{KL}, making it more difficult to differentiate normal knees from those exhibiting early osteoarthritic changes \cite{KL_1}. Such ambiguity often results in greater variability in interpretation among medical professionals, leading to potentially lower diagnostic confidence. Given these challenges, the early detection and precise classification of KL-2 are of significant clinical importance. As such, this study focuses on KL-0 healthy knee joints and KL-2 osteoarthritic knee joints as the primary research subjects denoted as $X_1$ and $X_2$, respectively.

\subsection{Employed knee database}
The Osteoarthritis Initiative (OAI) \cite{OAI} is a publicly available, large-scale, multi-centre longitudinal study designed to improve understanding of KOA. Established in 2002 with funding from the National Institutes of Health (NIH) and private industry partners, the OAI serves as a comprehensive resource for researchers seeking to advance the diagnosis, prevention, and treatment of KOA. The OAI includes a diverse cohort of 4,796 participants aged 45–79, with varying levels of risk for KOA. Over an extended follow-up period, participants undergo regular clinical assessments, imaging studies (e.g., X-rays and MRIs), and collection of biological samples, along with detailed surveys capturing lifestyle and health-related information.

\subsection{Data preprocessing}
\label{data_preprocessing}
The knee joint images were resized to 299 $\times$ 299 pixels, and image intensity distribution was normalized to the range [-1, 1]. After this, a total of 3,185 and 2,126 knee X-rays corresponding to KL-0 and KL-2, respectively, were retained. Due to the inherent class imbalance in the original dataset, the generated data also exhibited the same imbalance. To address this issue and mitigate its impact on accuracy when evaluating the effects of data augmentation across different models, bootstrapping-based oversampling \cite{nitesh2002smote} was applied to the KL-2 dataset. Moreover, since the labels of the two inputs ($X_1$ and $X_2$) must differ, a global permutation strategy was employed. Specifically, each KL-0 image was paired with every KL-2 image, creating distinct input pairs, resulting in a total of 3,318,240 input pairs. To simplify this, random samples of 50,000, 100,000, 500,000, and 1,000,000 input pairs $N$ were drawn from the total input pairs for experimentation, which will be discussed in Section \ref{samples_discussion}.

\subsection{Experimental details}
The model was initialized using the Kaiming procedure \cite{kaiming}. Adam optimizer \cite{adam} was employed for 1,000 epochs with a batch size of 32 and a learning rate of 1e-04. We implemented our approach using PyTorch v1.8.1 \cite{pytorch} on Nvidia TESLA A100 with 80 GB graphic memory.

\section{Results and discussion}
%The experimental findings will be shown and analysed in this section.

\subsection{Selection of the hyper-parameters in the hybrid loss}
\label{hyperparameters}
\begin{figure}[htbp]
\centering
\includegraphics[width=0.45\textwidth]{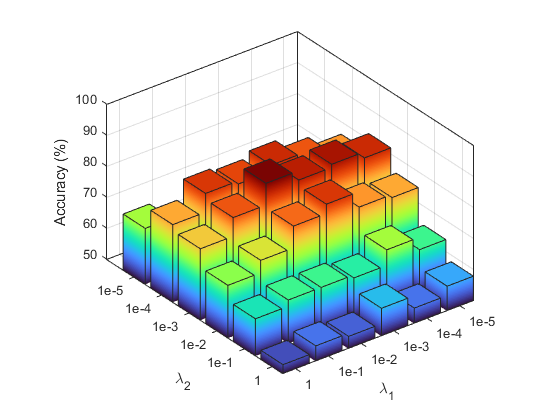}
\caption{Classification performance obtained using SVM-RBF with different values of $\lambda_1$ and $\lambda_2$ for key feature vector $h^K$.}
\label{3d}
\end{figure}

To evaluate the effectiveness of separating key and non-key features, we initially fixed the input pair size $N$ at 50,000 to reduce computational costs during hyper-parameter optimization. The classification performance of a Support Vector Machine with a Radial Basis Function (SVM-RBF) \cite{SVM} was utilized as a metric to assess the quality of extracted key features before applying data augmentation. The hyper-parameters $\lambda_1$ and $\lambda_2$ in the proposed hybrid loss strategy were tuned through a grid search over combinations of $\lambda_1, \lambda_2 \in [1e^{-5}, 1]$. Fig. \ref{3d} illustrates the variation in classification accuracy of the key feature vector $h^K$ as a function of different combinations of $\lambda_1$ and $\lambda_2$ under this fixed input pair configuration. As can be seen, $\lambda_2$ is more sensitive to accuracy compared to $\lambda_1$. Specifically, when $\lambda_2$ is set too large (i.e., between 1e-01 and 1), feature separation becomes almost ineffective, regardless of the value of $\lambda_1$. Conversely, when $\lambda_2$ lies within the range of 1e-05 to 1e-02, the accuracy varies with changes in $\lambda_1$. Among these combinations, the highest classification performance was achieved with $\lambda_1 = 1e^{-2}$ and $\lambda_2 = 1e^{-3}$, which were subsequently adopted as the optimal hyper-parameters for the following experiments.

\subsection{Visualization of the feature vectors}
\begin{figure}[htbp]
\centering
\subfloat[Original input $X_1$ (KL-0)]{
\label{X_1_present}
\begin{minipage}[t]{0.231\textwidth}
\centering
\includegraphics[width=1\textwidth]{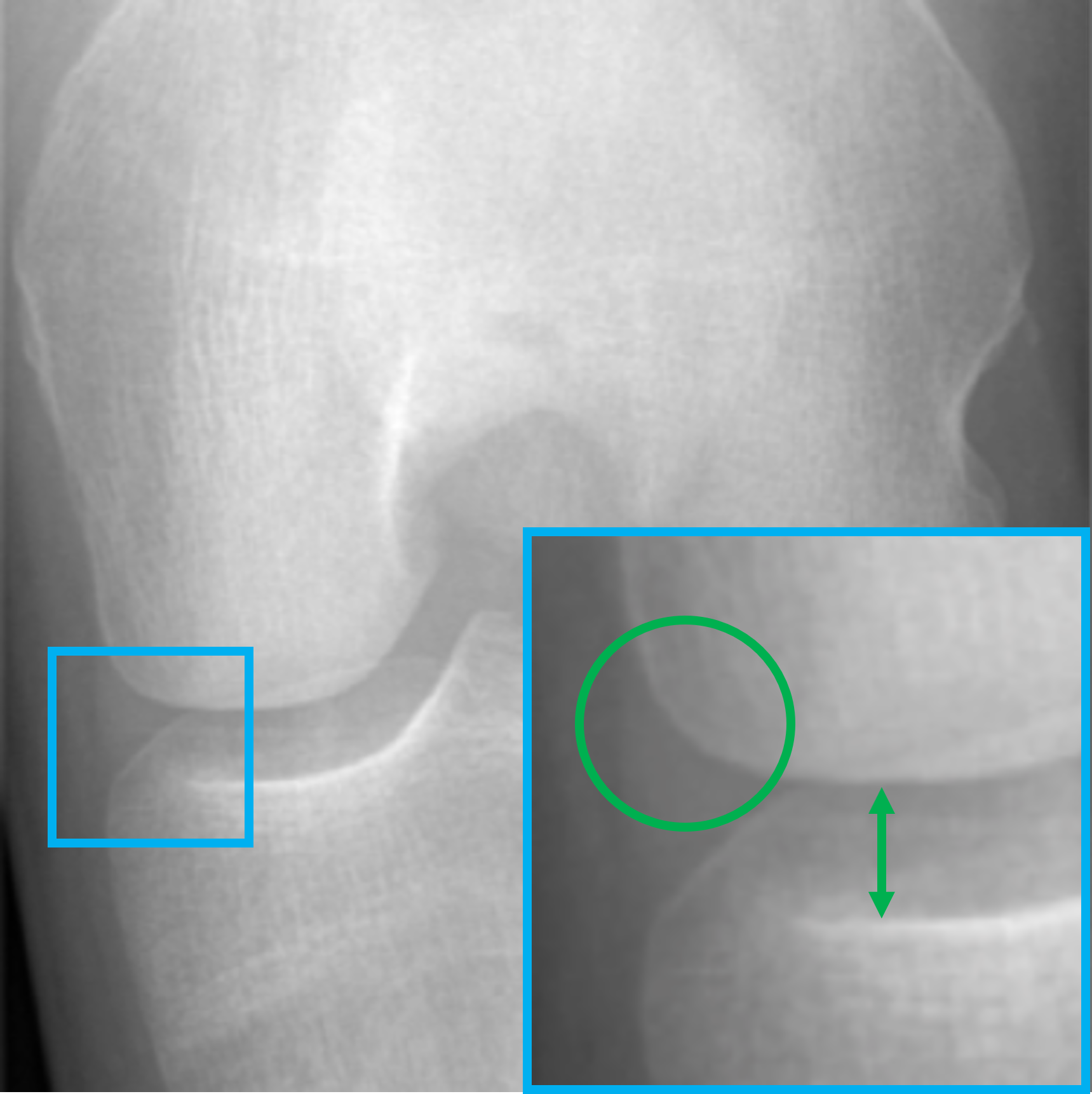}
\end{minipage}
}
\subfloat[key-exchanged output $X_1^\prime$ (KL-2)]{
\label{X_1_E_present}
\begin{minipage}[t]{0.231\textwidth}
\centering
\includegraphics[width=1\textwidth]{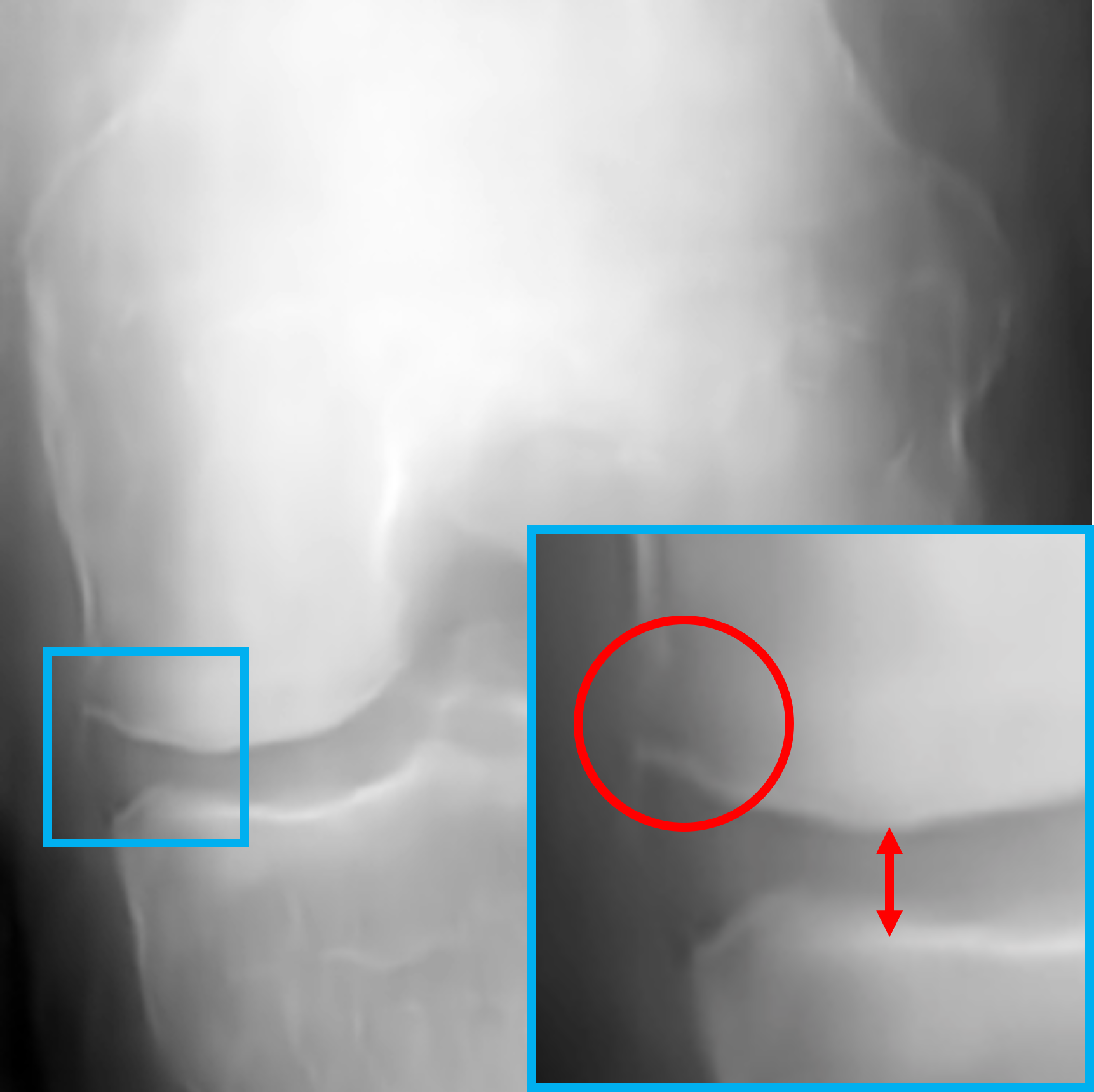}
\end{minipage}
}
\quad
\subfloat[Original input $X_2$ (KL-2)]{
\label{X_2_present}
\begin{minipage}[t]{0.231\textwidth}
\centering
\includegraphics[width=1\textwidth]{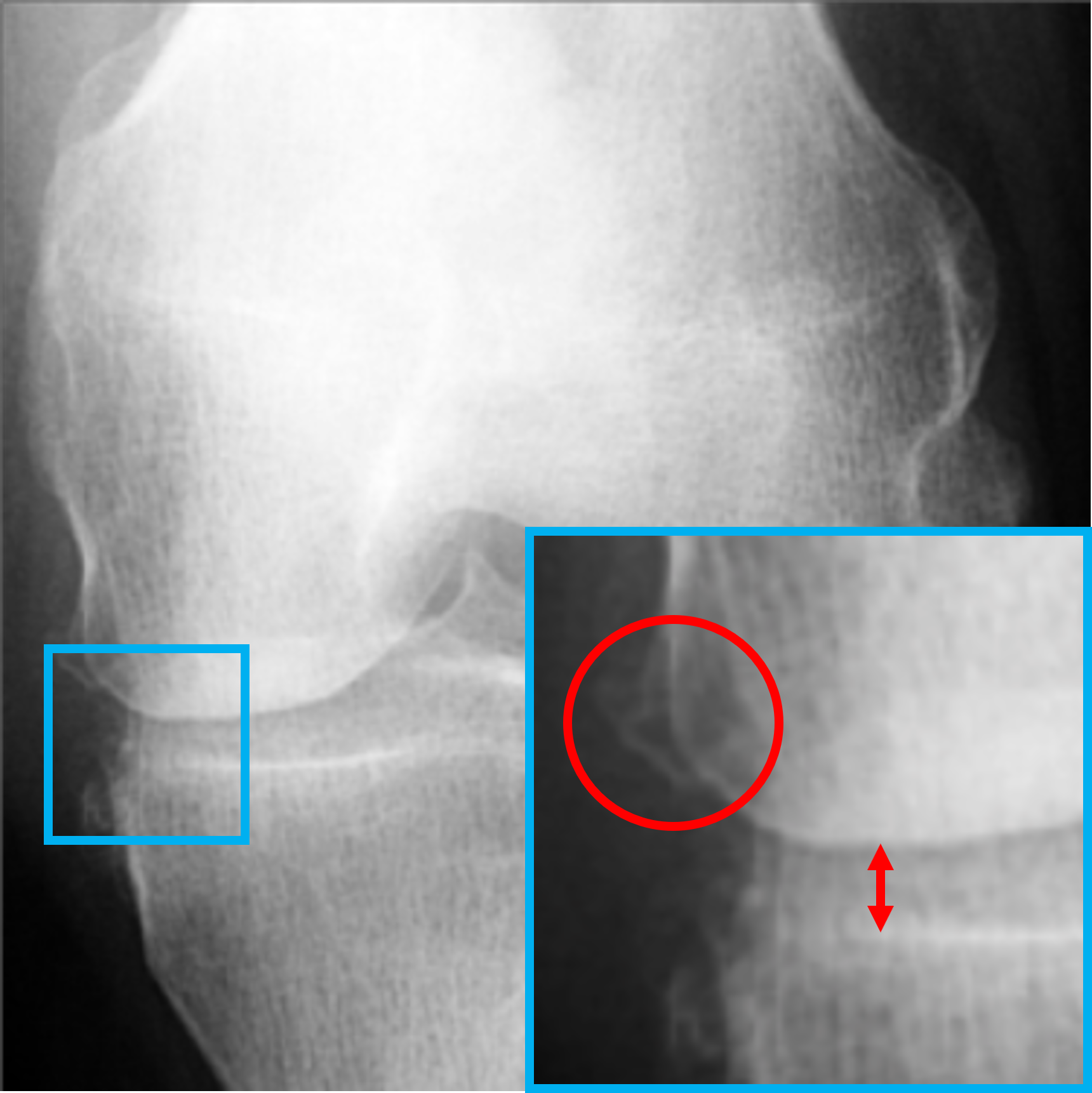}
\end{minipage}
}
\subfloat[Key-exchanged output $X_2^\prime$ (KL-0)]{
\label{X_2_E_present}
\begin{minipage}[t]{0.231\textwidth}
\centering
\includegraphics[width=1\textwidth]{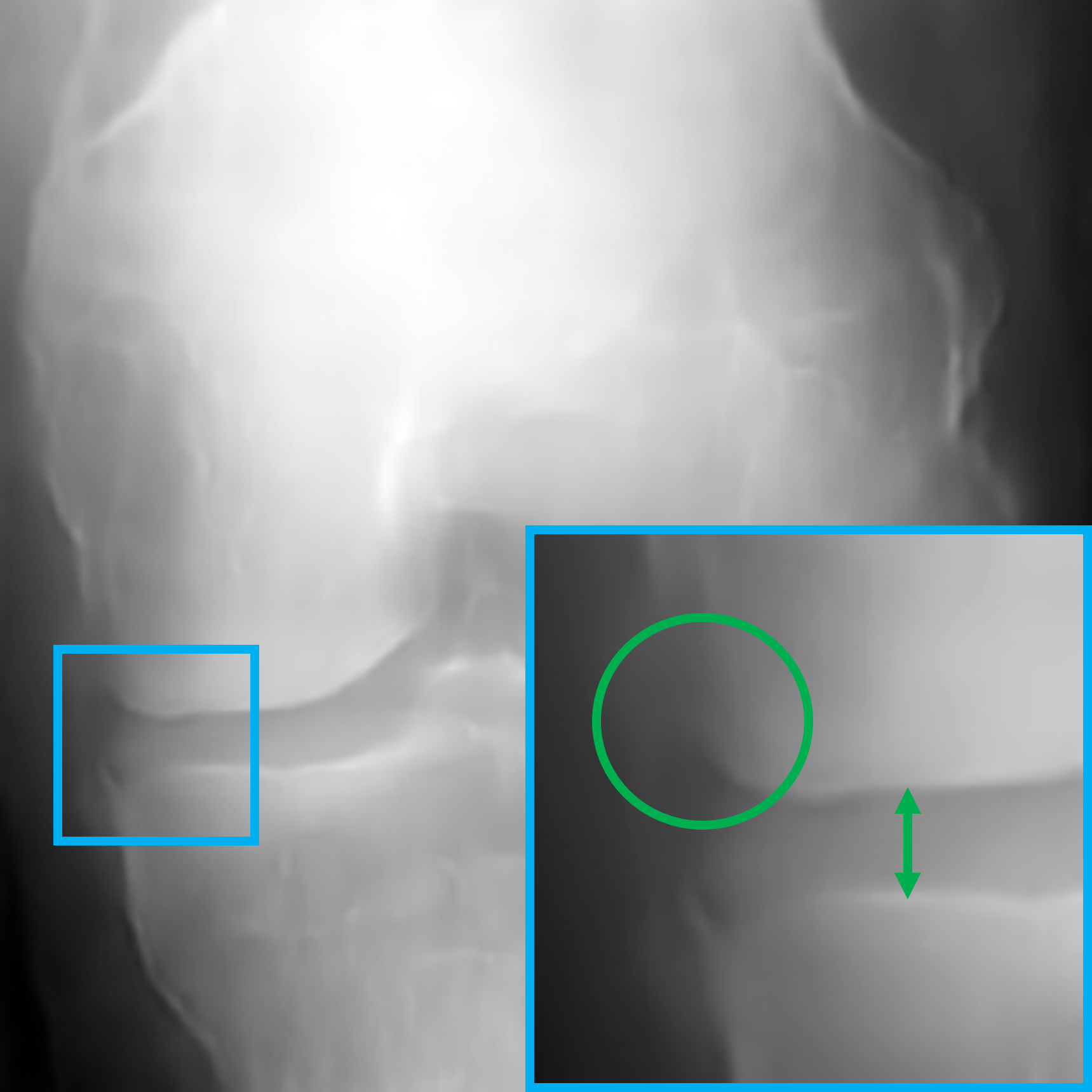}
\end{minipage}
}
\caption{Highlighted illustration of the original inputs and key-exchanged outputs. The green and red colours represent the absence and presence of symptoms for osteophytes and JSN, respectively. The circles and arrows represent the possible positions of the osteophytes and JSN modification, respectively.}
\label{Exchange_present}
\end{figure}

To provide a more intuitive understanding of the proposed key feature exchange mechanism, we visualized the original inputs and their key-exchanged counterparts, as shown in Fig. \ref{Exchange_present}. Specifically, $X_1$ (Fig. \ref{X_1_present}), a KL-0 healthy knee X-ray, is contrasted with its key-exchanged output $X_1^\prime$ (Fig. \ref{X_1_E_present}), while $X_2$ (Fig. \ref{X_2_present}), a KL-2 osteoarthritic knee X-ray, is compared with $X_2^\prime$ (Fig. \ref{X_2_E_present}). In $X^{\prime}_1$, pathological features associated with KL-2, JSN and osteophytes, are successfully introduced, while in $X^{\prime}_2$, these features are removed, restoring the joint space to resemble a healthy knee. It is noteworthy that, both $X^{\prime}_1$ and $X^{\prime}_2$ preserve the structural integrity and background details of the original images, demonstrating the model’s ability to selectively extract and modify key features without compromising realism. These visual results validate the effectiveness of our proposed global approach in separating key and non-key features, ensuring that the generated outputs are both diverse and valid. More details about the clinical validation will be discussed in Section \ref{clinical_validation}.

\subsection{Impact of the sampling size}
\label{samples_discussion}
\begin{figure}[htbp]
\label{tnse}
\centering
\subfloat[]{
\label{losses}
\begin{minipage}[t]{0.231\textwidth}
\centering
\includegraphics[width=1\textwidth]{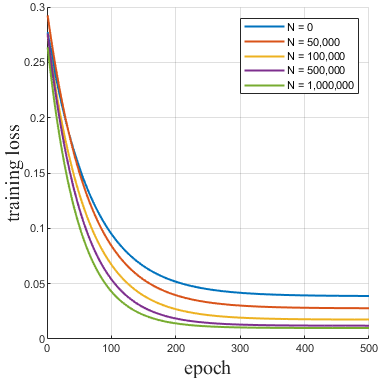}
\end{minipage}
}
\subfloat[]{
\label{accs}
\begin{minipage}[t]{0.231\textwidth}
\centering
\includegraphics[width=1\textwidth]{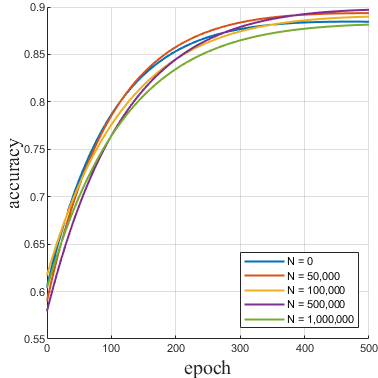}
\end{minipage}
}
\caption{Convergence (a) and performance (b) curves obtained using various sample sizes $N$ for 500 epochs.}
\label{samples_inputs}
\end{figure}

As presented in Section \ref{data_preprocessing}, we randomly sampled input pairs of sizes 50,000, 100,000, 500,000, and 1,000,000 from all possible input pairs. The discriminator used in this study is one of the State-Of-The-Art (SOTA) classifiers for KL-0 and KL-2 classification. Given its strong baseline performance, we first validated the effects of data augmentation on this classifier, as if the proposed method demonstrates significant improvements on this high-performing classifier, it is likely to yield even greater benefits for other models with lower initial performance. To evaluate the impact of the sampling size $N$ on the performance of data augmentation, for each value of $N$, five random samplings were performed. The discriminator was initialized and trained from scratch to assess both training and accuracy dynamics, and the average performance metrics were calculated. Fig. \ref{losses} shows that, for $N=0$ (i.e., using the original dataset without data augmentation), the training process required more epochs to converge, and the final training loss remained relatively high. With the introduction of augmented data ($N>0$), the convergence rate improved, and the final loss decreased significantly. This trend was most noticeable when $N$ increased from 0 to 50,000 and from 50,000 to 100,000. However, beyond $N=100,000$, the improvements in convergence speed and final loss became negligible, indicating diminishing returns from increasing the sample size further. On the other hand, Fig. \ref{accs} illustrates that, without data augmentation, the accuracy was limited by the inherent diversity of the original dataset. With $N=50,000$, the generated data provided additional diversity, leading to a noticeable improvement in accuracy. Further increases in $N$ to 100,000 yielded additional gains, but these gains were marginal compared to the improvements observed between $N=0$ and $N=50,000$. However, at $N=1,000,000$, the accuracy slightly decreased, likely due to the introduction of redundant or noisy data, which may have reduced the discriminator's ability to generalize.

Overall, larger sampling sizes (e.g., $N=1,000,000$) required significantly more computational resources and longer training times. On the other hand, smaller sampling sizes (e.g., $N=50,000$) did not fully exploit the benefits of data augmentation. The sampling size $N=100,000$ offered a balance between performance and computational efficiency, achieving robust gains in accuracy and convergence without the overhead associated with excessively large datasets. Based on these observations, we selected 100,000 as the optimal sample size for data augmentation in subsequent experiments.

\subsection{Effects of the proposed data augmentation}
\label{effects}
In this study, we evaluated the impact of the proposed approach as a data augmentation technique across various models selected from the literature. The augmented data ($X^{\prime}$) were combined with the original dataset ($X$) to form a larger training set $\mathcal{T^\prime}$, which was used to train several classification models. For convenience, we selected one high-performing model from each of the DenseNet, ResNet, and VGG series that achieved the best performance of the KOA classification in its group \cite{zhe}. Additionally, SOTA models proposed by Antony et al. \cite{antony1}, Tiulpin et al. \cite{tiuplin}, and Wang et al. \cite{zhe, wang2024transformer} were included for comparison. As shown in Table \ref{data_augmentation}, we evaluated two input cases: models trained solely on the original training set ($\mathcal{T}$) and those trained on the augmented training set ($\mathcal{T^\prime}$). The validation and test sets were kept fixed across all experiments to ensure a consistent evaluation of model performance. The results demonstrate that the key-exchanged outputs $X^{\prime}$ consistently improved the accuracy of all evaluated models, highlighting the effectiveness of the proposed augmentation strategy. Specifically, the VGG-11 model exhibited the largest accuracy improvement, achieving an increase of +1.98$\%$ (from 80.97$\%$ to 82.95$\%$), which can be attributed to its relatively lower baseline performance and its sensitivity to the enriched diversity provided by the augmented dataset. Similarly, DenseNet-201 and ResNet-18 showed accuracy gains of +1.21$\%$ and +1.63$\%$, respectively, further reinforcing the broad applicability of our approach. Among the SOTA models, Wang et al. \cite{wang2024transformer} achieved the highest overall accuracy (89.54$\%$), improving upon its already impressive baseline accuracy of 88.86$\%$. It is noteworthy that, while models with higher baseline accuracies experienced comparatively smaller improvements, this incremental gain is still noteworthy, given that these models are already operating near the upper limit of achievable accuracy. In addition to accuracy, improvements were also observed in precision and recall across most models. These findings highlight the proposed approach as a robust and effective data augmentation strategy, showcasing its ability to enhance model performance across diverse architectures and baseline accuracies.

\begin{table}[htbp]
\centering
\caption{Performance obtained using different training sets}
\setlength{\tabcolsep}{0.8mm}
\begin{tabular}{lcccc}  % 控制表格的格式
\toprule
Model &  Training set & Accuracy ($\%$) & Precision ($\%$) & Recall ($\%$)\\
\midrule
\multirow{2}{*}{DenseNet-201} & $\mathcal{T}$ &  84.43$\pm$1.27 &  81.95$\pm$1.55 & 81.04$\pm$1.39\\
&  $\mathcal{T^\prime}$ & 85.64$\pm$0.96 & 83.72$\pm$1.25 & 82.84$\pm$1.30\\
\midrule
\multirow{2}{*}{ResNet-18} & $\mathcal{T}$ &  83.59$\pm$1.58 &  80.42$\pm$1.91 & 79.35$\pm$1.77\\
& $\mathcal{T^\prime}$ & 85.22$\pm$1.19 & 82.81$\pm$1.39 & 81.74$\pm$1.33\\
\midrule
\multirow{2}{*}{VGG-11} & $\mathcal{T}$ &  80.97$\pm$2.07 &  77.86$\pm$2.33 & 75.76$\pm$2.01\\
&$\mathcal{T^\prime}$ & 82.95$\pm$1.54 & 80.12$\pm$1.79 & 78.08$\pm$1.68\\
\midrule
\multirow{2}{*}{Antony et al. \cite{antony1}} & $\mathcal{T}$ &  84.50$\pm$1.09 &  82.60$\pm$1.32 & 81.04$\pm$1.29\\
& $\mathcal{T^\prime}$ &  85.92$\pm$1.03 & 84.10$\pm$1.18 & 83.35$\pm$1.18\\
\midrule
\multirow{2}{*}{Tiulpin et al. \cite{tiuplin}} & $\mathcal{T}$ &  87.33$\pm$0.71 &  85.38$\pm$1.11 & 85.89$\pm$0.98\\
& $\mathcal{T^\prime}$ &  88.12$\pm$0.55 & 86.95$\pm$0.98 & 88.12$\pm$0.81\\
\midrule
\multirow{2}{*}{Wang et al. \cite{zhe}} & $\mathcal{T}$ &  88.38$\pm$0.44 &  87.69$\pm$0.88 & 88.85$\pm$0.91\\
& $\mathcal{T^\prime}$ &  89.11$\pm$0.35 & 89.14$\pm$0.77 & 90.22$\pm$0.71\\
\midrule
\multirow{2}{*}{Wang et al. \cite{wang2024transformer}} & $\mathcal{T}$ &  88.86$\pm$0.37 & 88.81$\pm$0.85 & 85.24$\pm$0.74\\
& $\mathcal{T^\prime}$ & 89.54$\pm$0.22 & 90.03$\pm$0.75 & 85.35$\pm$0.79\\
\bottomrule
\end{tabular}
\label{data_augmentation}
\end{table}

\subsection{Clinical validation for key-exchanged outputs}
\label{clinical_validation}
To evaluate the clinical validity of the key-exchanged outputs generated by the proposed model, we conducted a reader study involving three board-certified radiologists. The primary objective was to determine whether the generated outputs accurately and clinically reflected the intended key feature exchange, specifically whether the key-exchanged outputs aligned with their assigned post-exchange labels. A total of 100 randomly selected X-ray pairs, including original images ($X$) and their key-exchanged counterparts ($X^{\prime}$), were used for the study.  The evaluation was conducted in two phases. In the first phase, radiologists were presented with the key-exchanged outputs independently, without access to the original images. They were tasked with classifying each image as either KL-0 or KL-2 based on standard diagnostic criteria. In the second phase, radiologists reviewed the original images alongside their key-exchanged counterparts to assess whether the changes observed in the outputs were consistent with the expected key feature exchange. To quantify the outcomes, we computed three metrics: (1) Accuracy, defined as the percentage of key-exchanged outputs correctly classified by radiologists according to their assigned post-exchange labels; (2) Subjective Realism Score (SRS), where radiologists rated the anatomical plausibility and clinical realism of each key-exchanged output on a scale from 1 (poor) to 5 (excellent); and (3) Cohen's Kappa ($\kappa$) to measure inter-rater reliability among radiologists.

\begin{table}[htbp]
\caption{Clinical validation for key-exchanged outputs$^\star$}
\setlength{\tabcolsep}{4mm}
\centering
\label{tab:clinical_validation}
\begin{threeparttable}
\begin{tabular}{lccc}
\toprule
 & Accuracy (\%) & SRS   & $\kappa$\\
\midrule
Radiologist $\#$1 & 85 (82 / 88)  & 4.0 (3.9 / 4.1) & - \\
Radiologist $\#$2 & 88 (86 / 90)  & 4.2 (4.0 / 4.4)  & -\\
Radiologist $\#$3 & 86 (84 / 88)  & 4.2 (4.1 / 4.3) & - \\
\midrule
Average  & 86.33  & 4.13 & 0.78 \\
\bottomrule
\end{tabular}
\begin{tablenotes}
\footnotesize
\item[$\star$] For the Accuracy and SRS, the first value represents the average score, while the values in parentheses indicate the scores evaluated separately for outputs corresponding to the KL-0 and KL-2 labels (i.e., ($X_2^\prime$ / $X_1^\prime$)).
\end{tablenotes}
\end{threeparttable}
\end{table}

As can be seen in Table \ref{tab:clinical_validation}, the average accuracy across radiologists was 86.33\%, which indicates that the model effectively captures key feature exchanges in both healthy and osteoarthritic cases, with slightly higher accuracy for KL-2 outputs ($X_1^\prime$) compared to KL-0 outputs ($X_2^\prime$). This discrepancy may be attributed to the more distinct pathological features present in KL-2 cases, which are visually prominent and easier to identify. The SRS averaged 4.13 out of 5, radiologists noted that the changes in JSN and osteophyte presentation were realistic and consistent with clinical expectations, highlighting the anatomical plausibility and clinical realism of the generated outputs. Moreover, the inter-rater reliability, as measured by Cohen's Kappa (\(\kappa = 0.78\)), reflects substantial agreement among the radiologists, further supporting the consistency and reliability of the key-exchanged outputs, which suggests that the modifications introduced by the model were interpretable and clinically consistent across different evaluators. These findings demonstrate the clinical utility of the proposed model in generating augmented datasets that are both diagnostically meaningful and realistic.

\subsection{Discussion}
In this study, we proposed the Key-Exchange Convolutional Auto-Encoder (KECAE) as a novel data augmentation strategy to tackle the challenges posed by limited annotated data for KOA classification. By incorporating a key-exchange mechanism into the convolutional autoencoder architecture, our method enables the generation of synthetic data that not only enhances dataset diversity but also maintains a high degree of clinical validity. Furthermore, the experimental results demonstrated the robustness and adaptability of the KECAE-generated data across a wide range of classification models. The method consistently improved performance metrics such as accuracy, precision, and recall, regardless of the model architecture or its baseline accuracy, underscoring the effectiveness of the proposed data augmentation strategy, making it a valuable tool for advancing classification performance in KOA tasks. Several points are to be discussed.

\subsubsection{Key and non-key feature separation, how it works?}
Initially, the two feature vectors extracted from each input image by dual convolutional encoders are random and do not have clear meanings, resulting in a total of two pairs of feature vectors ($h_1^1, h_1^2$ and $h_2^1, h_2^2$). However, as the model trains, it progressively refines these vectors, separating them and assigning them meaningful roles. Specifically, to enhance the distinction between the two feature vectors ($h_1^1, h_1^2$) extracted from the input image $X_1$, LDA is employed to maximize the distance between the vectors while minimizing the variance within each. This algorithm is also applied to the pair of feature vectors ($h_2^1, h_2^2$) from another input $X_2$. Subsequently, one vector from each pair (e.g., $h_1^1$ from $X_1$ and $h_2^1$ from $X_2$) is selected and exchanged, forming two new pairs of vectors ($h_2^1, h_1^2$ and $h_1^1, h_2^2$). Each of these pairs undergoes element-wise addition to obtain two feature-exchanged vectors ($h_1^\prime, h_2^\prime$). Finally, an introduced discriminator supervises the two feature-exchanged images ($X_1^\prime, X_2^\prime$) decoded from these vectors, ensuring that their new labels correspond to the labels associated with the exchanged vectors. Specifically, the discriminator supervises the key-exchanged images to enforce that the labels of these new images correspond to the labels associated with the exchanged key features, which means that after exchanging the key features between images, the resulting images should be classified according to the labels of the key features. Through these two constraints, LDA and the supervision from the discriminator, as the model trains, the exchanged vectors are considered as key feature vectors ($h_1^K, h_2^K$), while the others are considered non-key feature vectors ($h_1^N, h_2^N$). In this study, key features correspond to the symptoms of KOA with KL-2 grade (JSN and osteophytes), while the non-key features represent the overall background of the knee joint and the parts with less informative content for classification.

\subsubsection{Strengths and limitations}
Our study offers several significant strengths. From a clinical perspective, the focus on early KOA diagnosis is particularly valuable, as timely identification can guide patients toward early physical interventions, potentially delaying the onset and progression of osteoarthritis. From a technical perspective, our key-exchange mechanism effectively captures and modifies key pathological and non-pathological features while preserving anatomical integrity. Therefore, our model generates augmented data that align with the distribution of the original dataset. Moreover, the adaptability of our data augmentation strategy across multiple classification architectures underscores its versatility and generalizability. Importantly, our augmentation approach does not overcompromise the realism or clinical validity of the images, as confirmed through clinical validation involving expert radiologists. On the other hand, this study has several limitations, primarily stemming from its reliance on data from the OAI database. While the OAI provides a well-annotated dataset, the generalizability of the proposed approach may be constrained when applied to data from other sources. Future research should leverage additional large-scale databases to validate the method further and ensure its robustness across diverse populations and imaging protocols. Another limitation is the computationally intensive and time-consuming nature of manually tuning the hyper-parameters in the hybrid loss strategy. Incorporating advanced optimization techniques, such as evolutionary algorithms, could streamline this process, significantly enhancing efficiency and reducing computational overhead.

\section{Statements}
The views and opinions presented in this manuscript, which employs OAI data, do not necessarily represent those of the OAI investigators, the NIH, or the private funding partners. 

\section{Acknowledgments}
The authors gratefully acknowledge the support of the French National Research Agency (ANR) through the ANR-20-CE45-0013-01 project. We also extend our sincere gratitude to the study participants, the dedicated clinical staff, and the coordinating centre at UCSF for their invaluable contributions.

\bibliographystyle{IEEEtran}
\bibliography{cas-refs}

\begin{thebibliography}{10}
\providecommand{\url}[1]{#1}
\csname url@samestyle\endcsname
\providecommand{\newblock}{\relax}
\providecommand{\bibinfo}[2]{#2}
\providecommand{\BIBentrySTDinterwordspacing}{\spaceskip=0pt\relax}
\providecommand{\BIBentryALTinterwordstretchfactor}{4}
\providecommand{\BIBentryALTinterwordspacing}{\spaceskip=\fontdimen2\font plus
\BIBentryALTinterwordstretchfactor\fontdimen3\font minus \fontdimen4\font\relax}
\providecommand{\BIBforeignlanguage}[2]{{%
\expandafter\ifx\csname l@#1\endcsname\relax
\typeout{** WARNING: IEEEtran.bst: No hyphenation pattern has been}%
\typeout{** loaded for the language `#1'. Using the pattern for}%
\typeout{** the default language instead.}%
\else
\language=\csname l@#1\endcsname
\fi
#2}}
\providecommand{\BIBdecl}{\relax}
\BIBdecl

\bibitem{kneeoa}
R.~F. Loeser, S.~R. Goldring, C.~R. Scanzello, and M.~B. Goldring, ``Osteoarthritis: a disease of the joint as an organ,'' \emph{Arthritis and rheumatism}, vol.~64, no.~6, p. 1697, 2012.

\bibitem{multi-factor}
A.~Litwic, M.~H. Edwards, E.~M. Dennison, and C.~Cooper, ``{Epidemiology and burden of osteoarthritis},'' \emph{British Medical Bulletin}, vol. 105, no.~1, pp. 185--199, 01 2013.

\bibitem{cardiovascular}
\BIBentryALTinterwordspacing
G.~Singh, J.~D. Miller, F.~H. Lee, D.~Pettitt, and M.~W. Russell, ``Prevalence of cardiovascular disease risk factors among us adults with self-reported osteoarthritis: data from the third national health and nutrition examination survey.'' \emph{Population}, vol.~7, p.~17, 2002. [Online]. Available: \url{http://europepmc.org/abstract/MED/12416788}
\BIBentrySTDinterwordspacing

\bibitem{notclear}
J.~C. Mora, R.~Przkora, and Y.~Cruz-Almeida, ``Knee osteoarthritis: pathophysiology and current treatment modalities,'' \emph{Journal of pain research}, vol.~11, p. 2189, 2018.

\bibitem{weightloss}
A.~E. Wluka, C.~B. Lombard, and F.~M. Cicuttini, ``Tackling obesity in knee osteoarthritis,'' \emph{Nature Reviews Rheumatology}, vol.~9, no.~4, pp. 225--235, 2013.

\bibitem{KL}
J.~H. Kellgren and J.~S. Lawrence, ``{Radiological Assessment of Osteo-Arthrosis},'' \emph{Annals of the Rheumatic Diseases}, vol.~16, no.~4, p. 494, 1957.

\bibitem{shamir}
L.~Shamir, S.~M. Ling, W.~W. Scott, A.~Bos, N.~Orlov, T.~J. Macura, D.~M. Eckley, L.~Ferrucci, and I.~G. Goldberg, ``Knee x-ray image analysis method for automated detection of osteoarthritis,'' \emph{IEEE Transactions on Biomedical Engineering}, vol.~56, no.~2, pp. 407--415, 2008.

\bibitem{tiuplin}
A.~Tiulpin, J.~Thevenot, E.~Rahtu, P.~Lehenkari, and S.~Saarakkala, ``{Automatic knee osteoarthritis diagnosis from plain radiographs: A deep learning-based approach},'' \emph{Scientific Reports}, 2018.

\bibitem{zhe}
Z.~Wang, A.~Chetouani, D.~Hans, E.~Lespessailles, and R.~Jennane, ``Siamese-gap network for early detection of knee osteoarthritis,'' in \emph{2022 IEEE 19th International Symposium on Biomedical Imaging (ISBI)}, 2022, pp. 1--4.

\bibitem{zhe_ViT}
Z.~Wang, A.~Chetouani, and R.~Jennane, ``Transformer with selective shuffled position embedding using roi-exchange strategy for early detection of knee osteoarthritis,'' \emph{arXiv preprint arXiv:2304.08364}, 2023.

\bibitem{chen}
P.~Chen, L.~Gao, X.~Shi, K.~Allen, and L.~Yang, ``{Fully automatic knee osteoarthritis severity grading using deep neural networks with a novel ordinal loss},'' \emph{Computerized Medical Imaging and Graphics}, vol.~75, pp. 84--92, 2019.

\bibitem{dataset}
J.~R{\"o}glin, K.~Ziegeler, J.~Kube, F.~K{\"o}nig, K.-G. Hermann, and S.~Ortmann, ``Improving classification results on a small medical dataset using a gan; an outlook for dealing with rare disease datasets,'' \emph{Frontiers in Computer Science}, p. 102, 2022.

\bibitem{heavily_rely}
C.~Shorten and T.~M. Khoshgoftaar, ``A survey on image data augmentation for deep learning,'' \emph{Journal of big data}, vol.~6, no.~1, pp. 1--48, 2019.

\bibitem{AE}
H.~Bourlard and Y.~Kamp, ``Auto-association by multilayer perceptrons and singular value decomposition,'' \emph{Biological cybernetics}, vol.~59, pp. 291--4, 02 1988.

\bibitem{GAN}
\BIBentryALTinterwordspacing
I.~Goodfellow, J.~Pouget-Abadie, M.~Mirza, B.~Xu, D.~Warde-Farley, S.~Ozair, A.~Courville, and Y.~Bengio, ``Generative adversarial nets,'' \emph{Advances in neural information processing systems}, vol.~27, 2014. [Online]. Available: \url{https://proceedings.neurips.cc/paper_files/paper/2014/file/5ca3e9b122f61f8f06494c97b1afccf3-Paper.pdf}
\BIBentrySTDinterwordspacing

\bibitem{khozeimeh2021combining}
F.~Khozeimeh, D.~Sharifrazi, N.~H. Izadi, J.~H. Joloudari, A.~Shoeibi, R.~Alizadehsani, J.~M. Gorriz, S.~Hussain, Z.~A. Sani, H.~Moosaei \emph{et~al.}, ``Combining a convolutional neural network with autoencoders to predict the survival chance of covid-19 patients,'' \emph{Scientific Reports}, vol.~11, no.~1, p. 15343, 2021.

\bibitem{cnn}
A.~Krizhevsky, I.~Sutskever, and G.~E. Hinton, ``Imagenet classification with deep convolutional neural networks,'' \emph{Advances in neural information processing systems}, vol.~25, pp. 1097--1105, 2012.

\bibitem{pesteie2019adaptive}
M.~Pesteie, P.~Abolmaesumi, and R.~N. Rohling, ``Adaptive augmentation of medical data using independently conditional variational auto-encoders,'' \emph{IEEE transactions on medical imaging}, vol.~38, no.~12, pp. 2807--2820, 2019.

\bibitem{kingma2013auto}
D.~P. Kingma, ``Auto-encoding variational bayes,'' \emph{arXiv preprint arXiv:1312.6114}, 2013.

\bibitem{gan_3d}
F.~H. K. d.~S. Tanaka and C.~Aranha, ``Data augmentation using gans,'' \emph{arXiv}, 2019.

\bibitem{DT}
A.~J. Myles, R.~N. Feudale, Y.~Liu, N.~A. Woody, and S.~D. Brown, ``An introduction to decision tree modeling,'' \emph{Journal of Chemometrics: A Journal of the Chemometrics Society}, vol.~18, no.~6, pp. 275--285, 2004.

\bibitem{frid_gan}
M.~Frid-Adar, I.~Diamant, E.~Klang, M.~Amitai, J.~Goldberger, and H.~Greenspan, ``Gan-based synthetic medical image augmentation for increased cnn performance in liver lesion classification,'' \emph{Neurocomputing}, vol. 321, pp. 321--331, 2018.

\bibitem{DCGAN}
A.~Radford, L.~Metz, and S.~Chintala, ``Unsupervised representation learning with deep convolutional generative adversarial networks,'' \emph{arXiv}, 2015.

\bibitem{NE}
\BIBentryALTinterwordspacing
F.~Farnia and A.~Ozdaglar, ``Do gans always have nash equilibria?'' in \emph{International Conference on Machine Learning}.\hskip 1em plus 0.5em minus 0.4em\relax PMLR, 2020, pp. 3029--3039. [Online]. Available: \url{https://proceedings.mlr.press/v119/farnia20a.html}
\BIBentrySTDinterwordspacing

\bibitem{fisher1}
S.~Mika, G.~Ratsch, J.~Weston, B.~Scholkopf, and K.-R. Mullers, ``Fisher discriminant analysis with kernels,'' in \emph{Neural networks for signal processing IX: Proceedings of the 1999 IEEE signal processing society workshop (cat. no. 98th8468)}.\hskip 1em plus 0.5em minus 0.4em\relax Ieee, 1999, pp. 41--48.

\bibitem{fisher2}
R.~A. Fisher, ``The use of multiple measurements in taxonomic problems,'' \emph{Annals of eugenics}, vol.~7, no.~2, pp. 179--188, 1936.

\bibitem{zeiler2010deconvolutional}
M.~D. Zeiler, D.~Krishnan, G.~W. Taylor, and R.~Fergus, ``Deconvolutional networks,'' in \emph{2010 IEEE Computer Society Conference on computer vision and pattern recognition}.\hskip 1em plus 0.5em minus 0.4em\relax IEEE, 2010, pp. 2528--2535.

\bibitem{wang2024transformer}
Z.~Wang, A.~Chetouani, M.~Jarraya, D.~Hans, and R.~Jennane, ``Transformer with selective shuffled position embedding and key-patch exchange strategy for early detection of knee osteoarthritis,'' \emph{Expert Systems with Applications}, vol. 255, p. 124614, 2024.

\bibitem{KL_1}
D.~Hart and T.~Spector, ``Kellgren \& lawrence grade 1 osteophytes in the knee—doubtful or definite?'' \emph{Osteoarthritis and cartilage}, vol.~11, no.~2, pp. 149--150, 2003.

\bibitem{OAI}
G.~Lester, ``{The Osteoarthritis Initiative: A NIH Public–Private Partnership},'' \emph{HSS Journal: The Musculoskeletal Journal of Hospital for Special Surgery}, vol.~8, no.~1, pp. 62--63, 2011.

\bibitem{nitesh2002smote}
V.~C. Nitesh, ``Smote: synthetic minority over-sampling technique,'' \emph{J Artif Intell Res}, vol.~16, no.~1, p. 321, 2002.

\bibitem{kaiming}
K.~He, X.~Zhang, S.~Ren, and J.~Sun, ``Delving deep into rectifiers: Surpassing human-level performance on imagenet classification,'' in \emph{Proceedings of the IEEE international conference on computer vision}, 2015, pp. 1026--1034.

\bibitem{adam}
D.~P. Kingma and J.~Ba, ``Adam: A method for stochastic optimization,'' \emph{CoRR}, vol. abs/1412.6980, 2015.

\bibitem{pytorch}
A.~Paszke, S.~Gross, F.~Massa, A.~Lerer, J.~Bradbury, G.~Chanan, T.~Killeen, Z.~Lin, N.~Gimelshein, L.~Antiga, A.~Desmaison, A.~Köpf, E.~Yang, Z.~DeVito, M.~Raison, A.~Tejani, S.~Chilamkurthy, B.~Steiner, L.~Fang, J.~Bai, and S.~Chintala, ``Pytorch: An imperative style, high-performance deep learning library,'' 2019.

\bibitem{SVM}
C.~J. Burges, ``A tutorial on support vector machines for pattern recognition,'' \emph{Data mining and knowledge discovery}, vol.~2, no.~2, pp. 121--167, 1998.

\bibitem{antony1}
J.~Antony, K.~McGuinness, K.~Moran, and N.~E. O'Connor, ``Automatic detection of knee joints and quantification of knee osteoarthritis severity using convolutional neural networks,'' in \emph{Machine Learning and Data Mining in Pattern Recognition}, P.~Perner, Ed.\hskip 1em plus 0.5em minus 0.4em\relax Cham: Springer International Publishing, 2017, pp. 376--390.

\end{thebibliography}

\end{document}